\documentclass[aps,prb,twocolumn,floatfix,footinbib,showpacs,superscriptaddress]{revtex4-1}
\usepackage[english]{babel}
\usepackage[utf8]{inputenc}
\usepackage{graphicx}
\usepackage{fancyhdr}
\usepackage[active]{srcltx}
\usepackage{setspace}
\usepackage{float}
\usepackage{amsmath}
\usepackage{amsfonts}
\usepackage{natbib}
\usepackage{color}
\usepackage[colorlinks=true,linkcolor=blue,pagecolor=blue,filecolor=blue,menucolor=blue,urlcolor=blue,citecolor=blue,anchorcolor=blue]{hyperref}

\begin{document}

\title{Spin transfer torque and anisotropic conductance in spin orbit coupled graphene }
\author{Morteza Salehi}
\affiliation{Department of Physics, Bu-Ali Sina University, Hamedan 65178, Iran}
\author{Razieh Beiranvand}
\affiliation{Physics group, Department of basic science, Ayatollah Boroujerdi university, Boroujerd, Iran}
\author{Mohammad Alidoust}
\affiliation{Department of Physics, Norwegian University of Science and Technology, N-7491 Trondheim, Norway}

\begin{abstract}
We theoretically study spin transfer torque (STT) in a graphene system with spin orbit coupling (SOC). We consider a graphene-based junction where the spin orbit coupled region is sandwiched between two ferromagnetic (F) segments. The magnetization in each ferromagnetic segment can possess arbitrary orientations. Our results show that the presence of SOC results in anisotropically modified STT, magnetoresistance, and charge conductance as a function of relative magnetization misalignment in the F regions. We have found that within the Klein regime, where particles hit the interfaces perpendicularly, the spin-polarized Dirac fermions transmit perfectly through the boundaries of a F-F junction (i.e., with zero reflection), regardless of the relative magnetization misalignment and exert zero STT. In the presence of SOC, however, due to band structure modification, a nonzero STT reappears. Our findings can be exploited for experimentally examining proximity-induced SOC into a graphene system.     
\end{abstract}

\date{\today}

\maketitle

\section{Introduction}

The spin polarization and spin polarized currents play key roles in spintronics devices such as quantum computers and random access memories.\cite{I.Zutic,S.A.Wolf,I.Zutic3} The orientation of spin polarization in a ferromagnetic layer can be manipulated by injecting a spin polarized current into the system. The spin polarized current exerts spin torque on local magnetic moments through spin-spin interactions. Therefore, the spin polarized currents can carry information coherently and deliver it to a targeted outlet magnetic moment through a so-called spin transfer torque (STT) mechanism. \cite{D.Wang,A.Dyrdat,K.-H.Ding,B.Zhou,J.Chen,F.Herling,C.-C.Lin2,N.Tombros,Z.P.Niu,C.-C.Lin,AGM} This phenomenon can serve as underlying block in ultralow-power spin switches and nonvolatile magnetic random access memories. \cite{stronics1,stronics2} 

A device made of a graphene sheet might be an excellent material platform for implementing the STT mechanism because of supporting long spin diffusion length 5-20$\mu$m, gate-tunable carrier density, and high carrier mobility. \cite{stronics5,stronics6,halt2} The presence of spin orbit interaction promises robust spin-dependent phenomena and the potential to create efficient spin field-effect transistors. However, the inherent spin orbit interaction of carbon atoms is negligible. \cite{exp_so1,exp_so2}

Various spin mediated phenomena such as magnetism\cite{N.Tombros}, superconductivity\cite{Heersche2007Nature,Heersche2007SSC,Zhang2008PRL}, and spin orbit coupling (SOC) can be imprinted into a graphene layer by the virtue of proximity effect \cite{prox1,prox2,prox3,prox4,prox5,prox6,prox7,prox8,prox9}. The spin orbit mediated interactions and flexibility of graphene in extrinsically acquiring various properties have attracted much attention, both experimentally and theoretically\cite{equalspin3,equalspin4,Beiranvand2019JMMM,Beiranvand2020SR,J.Azizi,rameshti,Chakraborty2007PRB,Dedkov2008PRL,Varykhalov2012PRL,avsar2014nat,vahid,ramin,sattari,K.Zollner,stt_halt,Q.-P.Wu,P.Jiang,C.-S.Huang,D.Sinha,S.Acharjee,A.Bouhlal,M.Vogl}, and facilitated striking achievements in utilizing spin-dependent phenomena in functional spin-logic devices. For example, spin-dependent tunable conductivity for a graphene hybrid structure that resulted in the creation of an all-electrical spin-field effect switch \cite{spin-f-effect1,spin-f-effect2}. The induction of local magnetic moment and exchange interaction can be achieved by doping or coupling with magnetic insulator atoms, such as ferromagnetic insulator yttrium iron garnet (YIG), with unfilled $d$ or $f$ shells, to the $\pi$ orbitals of carbon atoms without introducing detrimental effects on the excellent transport properties of a graphene layer. On the other hand, the insulating materials do not contribute to current carrying channels. To effectively tune the carrier density (through chemical potential) in a graphene heterostructure, one may use a thin (poly-)methyl methacrylate. \cite{exp_f1} The extrinsic spin orbit interaction in graphene couples the spin of charged carriers to their momentum through carbon atoms that reside on a honeycomb lattice arrangement. Also, spin orbit interaction can originate intrinsically from the coupling between moving spins and angular momentum of electrons. Nevertheless, unlike numerous studies so far, there is no consensus on the strength and type of induced SOC in graphene \cite{exp_f1,ex2,ex3,exp_so1,exp_so2,teor_stt_gr1,teor_stt_gr3,C.-C.Lin2}

\begin{figure}[t!]
\includegraphics[clip, trim=1.5cm 4.8cm 0.8cm 4.0cm, width=8.50cm,height=4.0cm]{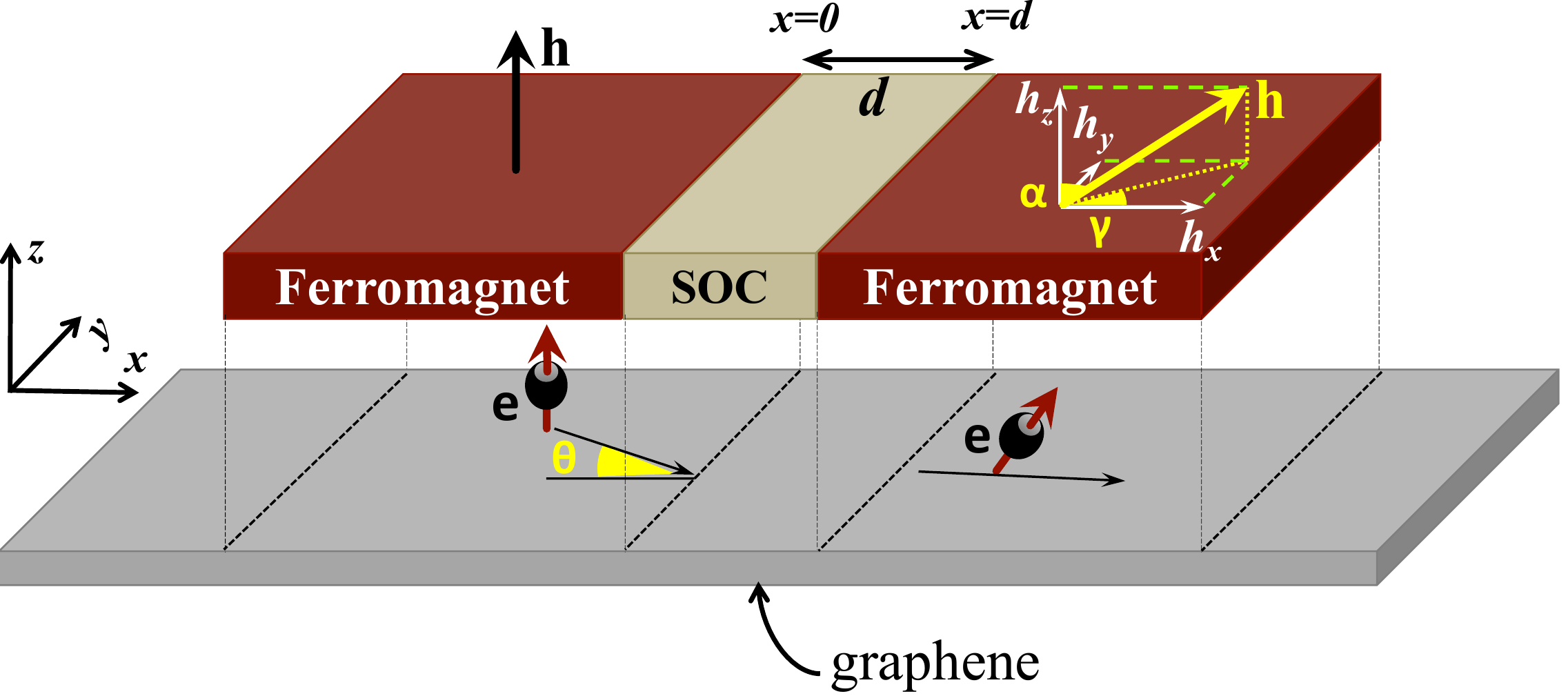}
\caption{\label{fig1} (Color online). Schematic set-up of the graphene-based junction. The system resides in the $xy$-plane and the interfaces are located at $x=0,d$. The magnetization in the right ferromagnetic region can have three components (i.e., an arbitrary orientation determined by spherical angles $\alpha, \gamma$) while magnetization in the left ferromagnetic region is oriented along the $z$ axis. Both magnetism and SOC are induced into the graphene layer through the virtue of proximity effect. By applying a voltage difference through the outmost left and right edges of the graphene layer in the $x$ direction, a charged particle, $e$, with its spin orientated along the $z$ axis hits the SOC interface with an angle $\theta$ with respect to the normal line perpendicular to the interface.}
\end{figure}

In this paper, we propose a spin valve configuration to unambiguously detect the presence of extrinsic SOC in a graphene layer. We consider a heterostructure of ferromagnet-SOC-ferromagnet (F-SOC-F) made of graphene where magnetization can possess an arbitrary orientations in the F regions. The schematic set-up is shown in Fig.~\ref{fig1}. Through the continuity equation, we derive generalized spin and charge currents and associated spin torque components in a ferromagnetic spin orbit coupled graphene layer. By making use of this generalized formalism and utilizing a wave function approach, we study STT, conductance, and magnetoresistance in the ferromagnetic spin valve, shown in Fig.~\ref{fig1}, in the presence and absence of SOC. Our results illustrate that the presence of SOC results in an anisotropically modified STT in response to the magentization misalignment in the ferromagnetic regions. This anisotropic behavior disappears in the absence of SOC. We find that similar isotropic and anisotropic behavior in the absence and presence of SOC, respectively, can be detected through magnetoresistance and charge conductance spectroscopy experiments. Also, we have considered a ferromagnet-ferromagnet configuration and obtained analytical expressions for the reflection and transmission coefficients. We find that in the Klein regime, the spin-polarized Dirac fermions transmit through the junction without any reflection, independent of the magnetization misalignment in the ferromagnetic regions. This perfect transmission results in zero STT in this regime. However, in the presence of SOC, this phenomenon disappears, i.e., the reflection coefficients and associated STT reappear. Our findings point to prominent signatures of induced SOC into a graphene layer within a spin-valve configuration.    

The paper is organized as follows. In Sec. \ref{formalism}, we summarize the formalism used and derive spin current components and associated STT. In Sec. \ref{results}, we utilize this formalism and study a spin-valve structure and present the results of STT, charge conductance, and magnetoresistance in the presence and absence of SOC. Finally, in Sec. \ref{conclusion}, we give concluding remarks.

\section{Theory and Formalism}\label{formalism}

The low-energy quasiparticles in a magnetized graphene sheet with extrinsic SOC are governed by the following effective Hamiltonian\cite{soc_gr1,soc_gr2}
\begin{subequations}\label{hamil}
\begin{align}
&\mathcal{H}{=}\int d \textbf{r}\Psi^\dagger(\textbf{r}) H(\textbf{r})\Psi(\textbf{r}),\\
&H(\textbf{r}){=}-i\hbar v_\text{F} s_0  \boldsymbol{\sigma}\cdot \boldsymbol{\nabla}+\lambda ({\bf s}\times\boldsymbol{\sigma})_z  +\textbf{h}\cdot{\bf s} \sigma_0-\mu.
\end{align}
\end{subequations}
Here, $\textbf{r}=(x,y,0)$ spans the two-dimensional plane of graphene, $v_\text{F}$ is the Fermi velocity, the strength of Rashba SOC is denoted by $\lambda$, magnetization $\textbf{h}=(h_x,h_y,h_z)$ possesses three components, and chemical potential $\mu$ can be tuned by an external electrode. The sublattice pseudospin (A,B) and real spin ($\uparrow,\downarrow$) degrees of freedom are shown by the $2$$\times$$ 2$ Pauli matrices; $\boldsymbol{\sigma}=(\sigma_x,\sigma_y,\sigma_z)$ and $\mathbf{s}=(s_x,s_y,s_z)$, respectively. The unity matrices are given by $\sigma_0$ and $s_0$. The four-component quasiparticle creation and annihilation operators $\Psi^\dagger(\textbf{r})=(\psi_{\text{A}\uparrow}^\dagger,\psi_{\text{B}\uparrow}^\dagger,\psi_{\text{A}\downarrow}^\dagger,\psi_{\text{B}\downarrow}^\dagger)$ and $\Psi(\textbf{r})=(\psi_{\text{A}\uparrow},\psi_{\text{B}\uparrow},\psi_{\text{A}\downarrow},\psi_{\text{B}\downarrow})^T$ obey the fermionic anticommutation relation. Throughout the paper, we have neglected the contribution of staggered on-site intrinsic SOC\cite{soc_gr1,soc_gr2}.

To obtain spin-polarized current and STT, a first-order time derivative of the components of spin-density, $\boldsymbol{{\cal S}}_j(\textbf{r})=\Psi^\dagger(\textbf{r})s_j\sigma_0\Psi(\textbf{r})$, is evaluated that results in the equation of motion
\begin{equation}
\partial_t \boldsymbol{{\cal S}}_j(\textbf{r})=\partial_t \Psi^\dagger(\textbf{r})\left\{{ s}_j \sigma_0\right\}\Psi(\textbf{r}) + \Psi^\dagger(\textbf{r})\left\{{ s}_j  \sigma_0\right\}\partial_t \Psi(\textbf{r}).
\label{Eq.DE01}
\end{equation}
We have defined index $j\equiv x,y,z$ for the spin components. By setting $\hbar=v_\text{F}=1$ and incorporating the fermionic anticommutation relation we find 
\begin{subequations}\label{Eq.FieldsCommutators}
 \begin{align}
& \left[\mathcal{H}, \Psi(\textbf{r})\right]=\nonumber\\
&\left\{ i s_0  \boldsymbol{\sigma}\cdot \boldsymbol{\nabla}-\textbf{h}\cdot{\bf s} \sigma_0 -\lambda (\textbf{s}\times \boldsymbol{\sigma})_z \right\}\Psi(\textbf{r}),\\
 & \left[\mathcal{H}, \Psi^\dagger(\textbf{r})\right]= \nonumber\\
 &i\boldsymbol{\nabla}\Psi^\dagger(\textbf{r})\left\{ s_0  \boldsymbol{\sigma}\right\}+\Psi^\dagger(\textbf{r}) \left\{\textbf{h}\cdot{\bf s} \sigma_0+\lambda (\textbf{s}\times \boldsymbol{\sigma})_z\right\}.
 \end{align}
 \end{subequations}
By substituting Eqs.~(\ref{Eq.FieldsCommutators}) into the equation of motion, Eq.~(\ref{Eq.DE01}), we obtain
 \begin{align}
\partial_t \boldsymbol{{\cal S}}_j(\textbf{r})=  &-\boldsymbol{\nabla}\Psi^\dagger(\textbf{r})\left\{ s_0  \boldsymbol{\sigma}\right\}\left\{s_j  \sigma_0\right\}\Psi(\textbf{r})\nonumber \\
& + i\Psi^\dagger(\textbf{r}) \left\{\textbf{h}\cdot\textbf{s} \sigma_0\right\}\left\{s_j  \sigma_0\right\}\Psi(\textbf{r}) \nonumber \\
&- \Psi^\dagger(\textbf{r})\left\{s_j  \sigma_0\right\}\left\{ s_0  \boldsymbol{\sigma}\right\} \boldsymbol{\nabla}\Psi^\dagger(\textbf{r})\nonumber\\
& - i\Psi^\dagger(\textbf{r})\left\{ s_j\sigma_0\right\}\left\{ \mathbf{h.s}  \sigma_0\right\}\Psi(\textbf{r}) \nonumber \\
&+ i \lambda \Psi^\dagger(\textbf{r})\left\{s_j\sigma_0\right\}(\textbf{s}\times \boldsymbol{\sigma})_z \Psi(\textbf{r}) \nonumber \\
&- i \lambda \Psi^\dagger(\textbf{r})(\textbf{s}\times \boldsymbol{\sigma})_z \left\{ s_j\sigma_0\right \} \Psi(\textbf{r}).
 \end{align}
This equation can be recast as
\begin{align}
\nonumber\partial_t \boldsymbol{{\cal S}}_j(\textbf{r})= & -\boldsymbol{\nabla}.\left(\Psi^\dagger(\mathbf{r})\left\{ s_j \boldsymbol{\sigma}\right\}\Psi(\mathbf{r})\right)  \nonumber \\ &+ 2 \Psi^\dagger(\textbf{r})(\textbf{h}\times \textbf{s})_j\Psi(\textbf{r}) \nonumber \\&+i \lambda \Psi^\dagger (\textbf{r})\left[s_j \sigma_0 , (\textbf{s}\times \boldsymbol{\sigma})_z\right]\Psi(\textbf{r}).
\label{Eq.CE01}
\end{align}
Considering Eq.~(\ref{Eq.CE01}), the spin-polarized current density $\boldsymbol{\mathcal{J}}^s_j$ and the total STT density $d\boldsymbol{\mathcal{T}}_j$ can be defined by 
\begin{subequations}\label{Eq.STTSpinCurrent}
 \begin{align}
 &\boldsymbol{\mathcal{J}}^s_j=\Psi^\dagger(\mathbf{r})\left\{ s_j \boldsymbol{\sigma}\right\}\Psi(\mathbf{r}),
\label{Eq.SpinCurrent}\\
&d\boldsymbol{T}_j=2\Psi^\dagger(\mathbf{r})(\textbf{h}\times \textbf{s})_j \Psi(\textbf{r}), \label{Eq.6b}\\
& d\boldsymbol{\mathfrak{T}}_j=i \lambda \Psi^\dagger (\textbf{r})\left[s_j \sigma_0 , (\textbf{s}\times \boldsymbol{\sigma})_z\right]\Psi(\textbf{r}),\label{Eq.6c}\\
& d\boldsymbol{\mathcal{T}}_j=d\boldsymbol{T}_j+d\boldsymbol{\mathfrak{T}}_j.
 \end{align}
 \end{subequations}
Using these definitions, the continuity equation reads
\begin{equation}
\partial_t \boldsymbol{{\cal S}}_j(\textbf{r})+\boldsymbol{\nabla}\cdot\boldsymbol{\mathcal{J}}^s_j= d\boldsymbol{\mathcal{T}}_j,
\end{equation} 
where $d\boldsymbol{\cal T}_j$ is the $j$ component of the STT density. In the absence of SOC, $\lambda=0$, the continuity equation reduces to
\begin{eqnarray}
 &\partial_t \boldsymbol{S}_j(\textbf{r})+\boldsymbol{\nabla}\cdot\boldsymbol{J}^s_j=d\boldsymbol{T}_{j}.
 \label{Eq.CE022}
 \end{eqnarray}
In the steady state, the spin wave density is time-independent, i.e., $\partial_t \boldsymbol{{\cal S}}_j(\textbf{r})=\partial_t \boldsymbol{S}_j(\textbf{r})=0$ and STT can be expressed via an integration over a region of interest
\begin{equation}\label{torque_tot}
\boldsymbol{\cal T}_{j}=\int d\textbf{r} \; d\boldsymbol{\cal T}_{j}= \int d\textbf{r} \boldsymbol{\nabla}\cdot\boldsymbol{\cal J}^s_j.
\end{equation}
The divergence theorem can be used to convert the latest integral into a surface integral, namely,
\begin{equation}
\boldsymbol{\cal T}_{j}= \int d\textbf{a}\cdot\boldsymbol{\cal J}^s_j.
\label{Eq.STT1}
\end{equation} 
This equation illustrates that STT within the region of interest can be obtained by subtracting the incoming and outgoing spin-polarized currents into the surface.

\section{results and discussions}\label{results}

We consider a graphene sheet placed in the $xy$ plane as displayed in Fig.~\ref{fig1}. Two ferromagnetic electrodes and a material with a strong SOC are deposited on the graphene layer in a junction configuration. The junction has a length of $d$ and the interfaces are located at $x=0,d$. The magnetization within the left region is directed along the $z$ axis, $\textbf{h}_l=(0,0,h)$, whereas the magnetization orientation in the right ferromagnetic region can be determined by the polar $\alpha$ and azimuthal $\gamma$ angles. Therefore, the orientation of magnetization in the right ferromagnetic region can be expressed by $\textbf{h}_r=h(\sin\alpha\cos\gamma,\sin\alpha\sin\gamma,\cos\alpha)$. To simulate and study the spin torque exerted on the junction, we consider a situation where a voltage bias drives particles from the left ferromagnetic region into the right ferromagnetic region. Therefore, a quasiparticle whose spin is oriented along the $z$ axis, described by $\phi^{l+}_\uparrow(r)$ spinor, is scattered upon hitting the interface from the left side of the junction at $x=0$. Due to the arbitrary orientation of magnetization in the right ferromagnetic region, this quasiparticle can be reflected with the same $r_{\uparrow \uparrow}$ or rotated $r_{\uparrow\downarrow}$ spin orientation. Also, this quasiparticle can be transmitted into the right ferromagnetic region with either the same $t_{\uparrow\uparrow}$ or rotated spin orientation $t_{\uparrow\downarrow}$. The wave functions of the back scattered and transmitted quasiparticles with spin-up and spin-down states are denoted by $\phi^{l-}_\uparrow(r)$, $\phi^{l-}_\downarrow(r)$, $\phi^{r+}_\uparrow(r)$, and $\phi^{r+}_\downarrow(r)$, respectively. The directions of quasiparticle's motion, along and opposite to the direction of the $x$ axis, are marked by $\pm$, respectively. Therefore, the total wave function in the left ferromagentic region, $\Psi^l(\textbf{r})$, can be expressed by
\begin{equation}\label{tot_wv_l}
\Psi^l(\textbf{r})=\phi^{l+}_\uparrow(\textbf{r})+r_{\uparrow\uparrow}\phi^{l-}_\uparrow(\textbf{r})+r_{\uparrow\downarrow}\phi^{l-}_\downarrow(\textbf{r}).
\end{equation}
These wave functions can be obtained by setting $\lambda=0$ and diagonalizing Hamiltonian (\ref{hamil})
\begin{subequations}
\begin{align}
&\phi^{l\pm}_\uparrow(\textbf{r})=\frac{1}{\sqrt{2\cos\theta^l_\uparrow}}\left( 
\begin{array}{cccc}
1, & \pm e^{\pm i \theta^l_\uparrow}, & 0, & 0 
\end{array}
\right)^T e^{\pm i \textbf{k}^l_\uparrow\cdot\textbf{r}}, \\
&\phi^{l\pm}_\downarrow(\textbf{r})=\frac{1}{\sqrt{2\cos\theta^l_\downarrow}}\left( 
\begin{array}{cccc}
0, & 0, & -1, &  \mp e^{\pm i \theta^l_\downarrow} 
\end{array}
\right)^T e^{\pm i \textbf{k}^l_\downarrow\cdot\textbf{r}}.
\end{align}
\end{subequations}
Here $\theta^l_{\uparrow,\downarrow}$ are the propagation angles of spin-up and spin-down quasiparticles with respect to the normal unit vector perpendicular to the interface (see Fig.~\ref{fig1}). The wave vector of the spin-up and spin-down particles are given by
\begin{equation}
\textbf{k}^l_{\uparrow,\downarrow}=\frac{\epsilon+\mu\pm h}{\hbar v_F}.
\end{equation}
Likewise, the total wave function in the right ferromagnetic region can be expressed by
\begin{equation}\label{tot_wv_r}
\Psi^r(\textbf{r})=t_{\uparrow\uparrow}\phi^{r+}_\uparrow(\textbf{r})+t_{\uparrow\downarrow}\phi^{r+}_\downarrow(\textbf{r}).
\end{equation}
The spinors in the right ferromagnetic region read
\begin{subequations}
\begin{align}
&&\phi^{r+}_\uparrow(\textbf{r})=\frac{1}{\sqrt{2\cos\theta^r_\uparrow}}\left(\begin{array}{l}
+\cos(\alpha/2) \\ 
+\cos(\alpha/2)e^{i \theta^r_\uparrow}\\+ \sin(\alpha/2)e^{i\gamma} \\
+\sin(\alpha/2)e^{i \gamma+i\theta^r_\uparrow}
\end{array}\right)e^{+i \textbf{k}^r_\uparrow\cdot\textbf{r}},\\
&&\phi^{r+}_\downarrow(\textbf{r})=\frac{1}{\sqrt{2\cos\theta^r_\downarrow}}\left(\begin{array}{l}
+\sin(\alpha/2) \\ 
+\sin(\alpha/2)e^{i \theta^r_\downarrow}\\ -\cos(\alpha/2)e^{i\gamma} \\
-\cos(\alpha/2)e^{i \gamma+i\theta^r_\downarrow}
\end{array}\right)e^{+i \textbf{k}^r_\downarrow\cdot\textbf{r}},
\label{Eq.RightWaveFunction}
\end{align}
\end{subequations}
in which the wave vectors of the spin-up and spin-down quasiparticles are
\begin{equation}
\textbf{k}^r_{\uparrow,\downarrow}=\frac{\epsilon+\mu\pm h}{\hbar v_F}.
\end{equation}
The total wave function $\Psi^m(\textbf{r})$ in the middle region with a finite SOC is a linear combination of the wave functions of right-moving $\phi^{m+}_{\pm}$ and left-moving $\phi^{m-}_{\pm}$ quasiparticles. This combination can be expressed through four unknowns $a_{1,2,3,4}$ as
\begin{equation}\label{tot_wv_m}
\Psi^m(\textbf{r})= a_1 \phi^{m+}_{+}(\textbf{r})+a_2 \phi^{m-}_{+}(\textbf{r})
 + a_3 \phi^{m+}_{-}(\textbf{r})+a_4 \phi^{m-}_{-}(\textbf{r}).
\end{equation}
These spinors can be obtained by setting $\textbf{h}=0$ and diagonalizing Hamiltonian (\ref{hamil}) \cite{equalspin3,equalspin4,Beiranvand2019JMMM,Beiranvand2020SR},
\begin{subequations}
\begin{align}
\phi^{m\pm}_{+}(\textbf{r})=\frac{1}{\sqrt{4 f_+\cos\theta_+^m}}\left(\begin{array}{c}
\mp i f_+ e^{\mp  i \theta_+^m} \\ 
-i \\
+1\\
\pm f_+ e^{\pm i \theta_+^m}
\end{array}\right)e^{\pm i \textbf{k}^m_+\cdot\textbf{r}},\\
\phi^{m\pm}_{- }(\textbf{r})=\frac{1}{\sqrt{4 f_- \cos\theta_-^m}}\left(\begin{array}{c}
\pm f_- e^{\mp i \theta_-^m} \\ 
+1\\
 -i  \\
\mp f_- e^{\pm i \theta_-^m}
\end{array}\right)e^{\pm i \textbf{k}^m_-\cdot\textbf{r}}.
\label{Eq.RightWaveFunction}
\end{align}
\end{subequations}
In the middle region, the wave vectors are
\begin{subequations}
\begin{align}
&\textbf{k}^m_{\pm}=\frac{1}{\hbar v_F}\sqrt{(\epsilon+\mu)(\epsilon+\mu\pm \lambda)},\\
&f_\pm=\sqrt{1\pm 2 \lambda (\mu+\epsilon)^{-1}}.
\end{align}
\end{subequations}
To find the reflection and transmission probabilities, the total wave functions, introduced earlier within each region, should be matched at the interfaces, namely \cite{Beenakker}, 
\begin{subequations}
\begin{align}
\Psi^l(\textbf{r})&=\Psi^m(\textbf{r})|_{x=0}, \\
 \Psi^m(\textbf{r})&=\Psi^r(\textbf{r})|_{x=d}.
\label{Eq.FFBoundaryCondition}
\end{align}
\end{subequations}

According to Eq.~(\ref{torque_tot}), the total torque exerted on the junction due to spin-polarized current $\boldsymbol{\cal J}^s$ can be evaluated by obtaining the incoming and outgoing spin-polarized currents into surface $\textbf{a}$, surrounding the junction interface. In order to find the incoming and outgoing spin-polarized currents, we substitute the total wave functions (\ref{tot_wv_r}) and (\ref{tot_wv_l}) into Eq.~(\ref{Eq.SpinCurrent}). The resultant expressions for the components of the spin-polarized currents are summarized in Eqs.(\ref{jx})-(\ref{jz}) 

\begin{figure*}[t!]
\includegraphics[width=0.9\textwidth]{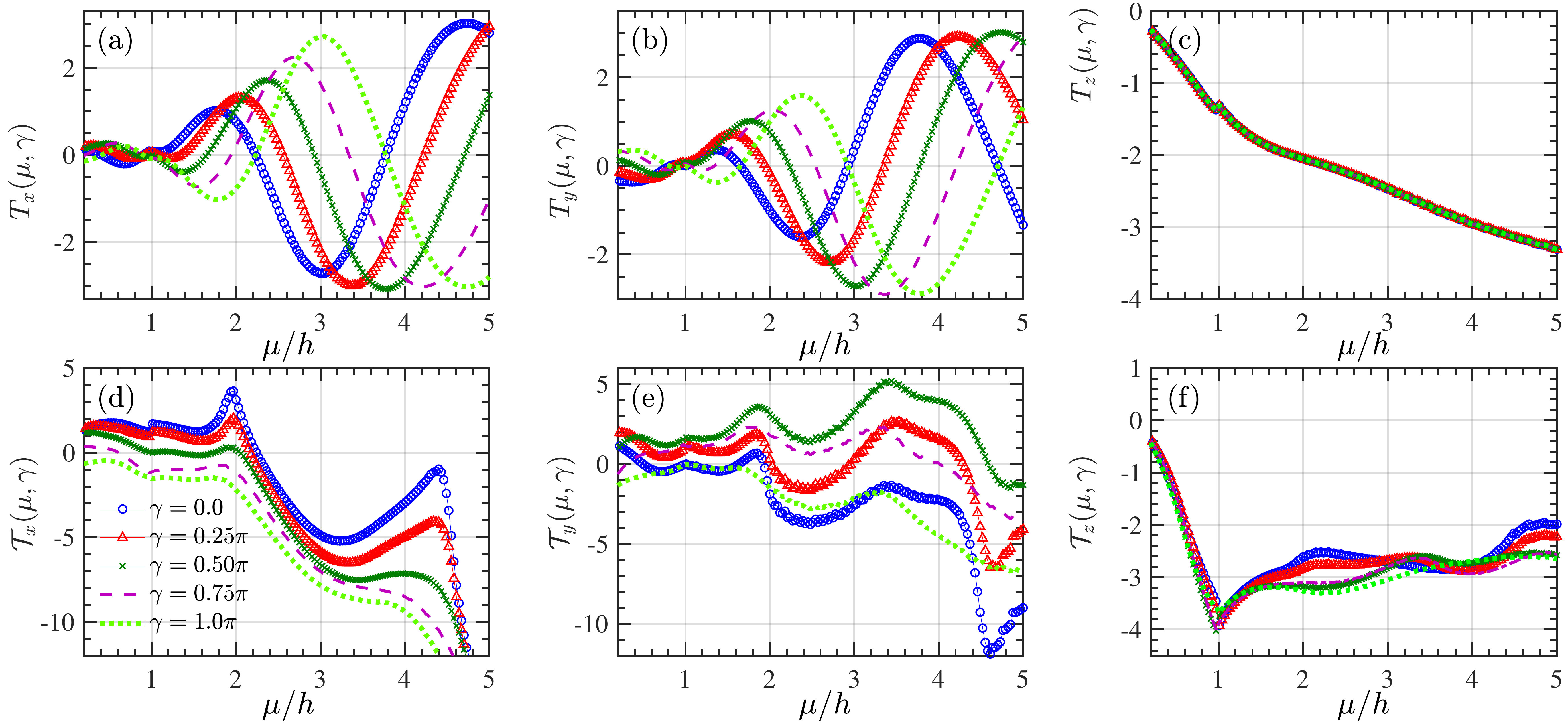}
\caption{\label{stt_mu} (Color online). The STT as a function of chemical potential $\mu$ for different values of $\gamma$. In (a)-(c) SOC is absent whereas in (d)-(e) SOC has a representative finite value $\lambda=1.5h$. The left, middle, and right columns show the $x$, $y$, and $z$ components of spin torque. The junction length and polar angle of magnetization in the right ferromagnetic region are set to $d=2\ell_F$ and $\alpha=0.75\pi$, respectively. }
\end{figure*}

\begin{widetext}
\begin{subequations}\label{jx}
\begin{align}
\boldsymbol{\cal J}^{ \text{s,l}}_{x} = \frac{\rho_{\uparrow}}{2}\left(
r_{\uparrow\downarrow}\langle \phi^{l,+}_\uparrow|s_x \sigma_x | \phi^{l,-}_\downarrow\rangle
+ r^*_{\uparrow\uparrow}r_{\uparrow\downarrow}\langle\phi^{l,-}_\uparrow|s_x \sigma_x|\phi^{l,-}_\downarrow\rangle 
\right)+\nonumber\\\frac{\rho_{\downarrow}}{2}\left( 
r_{\downarrow\uparrow }\langle \phi^{l,+}_\downarrow|s_x\sigma_x|\phi^{l,-}_\uparrow \rangle 
+  r^*_{\downarrow\downarrow} r_{\downarrow\uparrow} \langle \phi^{l,-}_\downarrow|s_x\sigma_x|\phi^{l,-}_\uparrow\rangle  
\right)+ \text{h.c.}~,
\end{align}
\begin{align}
\boldsymbol{\cal J}^{\text{s,r}}_{x}=\frac{\rho_{\uparrow}}{2}\left(
(|t_{\uparrow\uparrow}|^2-|t_{\uparrow\downarrow}|^2)\sin\alpha\cos\gamma
 +t^*_{\uparrow\uparrow}t_{\uparrow\downarrow} \langle \phi^{r +}_\uparrow| s_x \sigma_x | \phi^{r +}_\downarrow \rangle
 \right)+\nonumber\\\frac{\rho_{\downarrow}}{2}\left(
(|t_{\downarrow\uparrow}|^2-|t_{\downarrow\downarrow}|^2)\sin\alpha\cos\gamma
 +t^*_{\downarrow\downarrow}t_{\downarrow\uparrow} \langle \phi^{r +}_\downarrow| s_x \sigma_x | \phi^{r +}_\uparrow \rangle
 \right)+ \text{h.c.}~,
\end{align}
\end{subequations}
\begin{subequations}\label{jy}
\begin{align}
\boldsymbol{\cal J}^{ \text{s,l}}_{y} = \frac{\rho_{\uparrow}}{2}\left(
r_{\uparrow\downarrow}\langle \phi^{l,+}_\uparrow|s_y \sigma_x | \phi^{l,-}_\downarrow\rangle
 + r^*_{\uparrow\uparrow}r_{\uparrow\downarrow}\langle\phi^{l,-}_\uparrow|s_y \sigma_x|\phi^{l,-}_\downarrow\rangle 
\right)+\nonumber\\\frac{\rho_{\downarrow}}{2}\left( 
r_{\downarrow\uparrow }\langle \phi^{l,+}_\downarrow|s_y\sigma_x|\phi^{l,-}_\uparrow \rangle 
+  r^*_{\downarrow\downarrow} r_{\downarrow\uparrow} \langle \phi^{l,-}_\downarrow|s_y\sigma_x|\phi^{l,-}_\uparrow\rangle 
\right)+ \text{h.c.}~,
\end{align}
\begin{align}
\boldsymbol{\cal J}^{\text{s,r}}_{y}=\frac{\rho_{\uparrow}}{2}\left(
(|t_{\uparrow\uparrow}|^2-|t_{\uparrow\downarrow}|^2)\sin\alpha\sin\gamma 
 +t^*_{\uparrow\uparrow}t_{\uparrow\downarrow} \langle \phi^{r +}_\uparrow| s_y \sigma_x | \phi^{r +}_\downarrow \rangle
\right)+\nonumber\\\frac{\rho_{\downarrow}}{2}\left(
(|t_{\downarrow\uparrow}|^2-|t_{\downarrow\downarrow}|^2)\sin\alpha\sin\gamma
 +t^*_{\downarrow\downarrow}t_{\downarrow\uparrow} \langle \phi^{r +}_\downarrow| s_y \sigma_x | \phi^{r +}_\uparrow \rangle \right)+ \text{h.c.},
\end{align}
\end{subequations}
\begin{subequations}\label{jz}
\begin{equation}
\boldsymbol{\cal J}^{\text{s,l}}_{z}=\rho_{\uparrow}(1-|r_{\uparrow\uparrow}|^2+|r_{\uparrow\downarrow}|^2)+\rho_{\downarrow}(-1+|r_{\downarrow\downarrow}|^2-|r_{\downarrow\uparrow}|^2)~,
\end{equation}
\begin{align}
\boldsymbol{\cal J}^{ \text{s,r}}_{z}=
\frac{1}{2}\rho_{\uparrow}\left(
(|t_{\uparrow\uparrow}|^2-|t_{\uparrow\downarrow}|^2)\cos\alpha
+t^*_{\uparrow\uparrow}t_{\uparrow\downarrow} \langle \phi^{r +}_\uparrow| s_z \sigma_x | \phi^{r +}_\downarrow \rangle
\right)+ \nonumber \\ \frac{1}{2}\rho_{\downarrow}\left(
(|t_{\downarrow\uparrow}|^2-|t_{\downarrow\downarrow}|^2)\cos\alpha
 + t^*_{\downarrow\downarrow}t_{\downarrow\uparrow} \langle \phi^{r +}_\downarrow| s_z \sigma_x | \phi^{r +}_\uparrow \rangle
+ \text{h.c.}\right)~,
\end{align}
\end{subequations}
\end{widetext}
where $\rho_{\uparrow(\downarrow)}=2We^2|\epsilon+\mu\pm h|(h \pi \hbar v_F)^{-1}$ is the spin-polarized charge density in a junction with width $W$.
As can be seen, the obtained expressions for the components of the spin-polarized current are highly complicated. Moreover, the reflection and transmission probabilities in the presence of SOC are lengthy expressions that can be evaluated numerically only. Therefore, we omit presenting these expressions and directly discuss our numerical results.  
In the numerical study that follows, we have normalized energies by the induced exchange energy into the graphene layer, $h$, and lengths by $\ell_F=\hbar v_F/h$. If we consider a situation where the induced magnetization into the graphene layer is on the order of 0.1~meV, the characteristic length is on the order of $\ell_F\approx 100$~nm.

To begin, we have considered a junction configuration where the width of SOC region in Fig.~\ref{fig1} is zero, i.e., $d=0$. Therefore, the system reduces to a ferromagnet-ferromagnet junction with misaligned magnetization orientations. In this case, the reflection and transmission coefficients simplify and allow for achieving deeper insight into the scattering process.
To illustrate the influence of magnetization misalignment, we assume that the magnitude of magnetization in both sides of the junction is equal to $ h$, so the propagation directions obey $\theta^l_\uparrow=\theta^r_\uparrow=\theta_\uparrow$ and $\theta^l_\downarrow=\theta^r_\downarrow=\theta_\downarrow$. Performing the calculations, we arrive at these coefficients 
\begin{subequations}
\begin{align}
r_{\uparrow\uparrow}=& -\frac{2i}{D} e^{i \theta_\uparrow}(\sin \theta_\downarrow-\sin\theta_\uparrow )\sin^2\frac{\alpha}{2},\\
r_{\uparrow\downarrow}=& +\frac{e^{i\gamma}}{D}   (e^{i\theta_\uparrow} - e^{i\theta_\downarrow}) \sqrt{\cos\theta_\downarrow \cos\theta_\uparrow}\sin\alpha,\\
t_{\uparrow\uparrow}=& +\frac{4}{D}\cos\theta_\downarrow \cos\theta_\uparrow\cos\frac{\alpha}{2},\\
t_{\uparrow\downarrow}=& +\frac{4 e^{i\gamma}}{D}e^{i\frac{\theta_\uparrow-\theta_\downarrow}{2}}\sqrt{\cos\theta_\downarrow\cos\theta_\uparrow}\cos\frac{\theta_\uparrow+\theta_\downarrow}{2}\sin\frac{\alpha}{2},
\end{align}
\label{Eq.FF.RandT}
\end{subequations}
where the denominator, $D$, is given by
\begin{equation}
D= 1+\cos\alpha(\cos(\theta_\downarrow-\theta_\uparrow)-1)+3\cos\theta_\downarrow\cos\theta_\uparrow-\sin\theta_\downarrow\sin\theta_\uparrow.
\end{equation}
In the above expression, the first and second indices indicate the spin direction of the incoming particles and outgoing particles, respectively. Using this notation, the transport coefficients of spin-down incoming fermions read
\begin{figure*}[t!]
\includegraphics[clip, trim=1.5cm 7.8cm 0.4cm 2.0cm, width=0.9\textwidth]{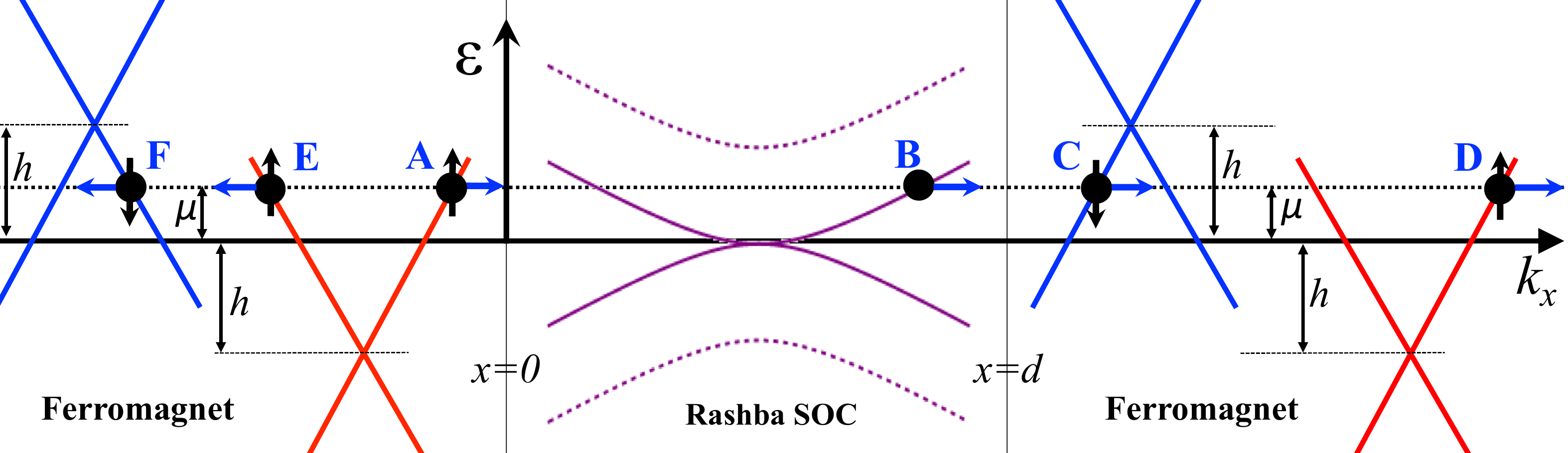}
\caption{\label{bs} (Color online). The band structure illustration of the scattering process. The vertical and horizontal axes are energy ($\epsilon$) and momentum ($k_x$), respectively. The ferromagnetic regions are described by the spin-up (red) and spin-down (blue) subbands. The two subbands are separated by amount $h$ in energy so that when $h=0$, the two subbands merge into a single Dirac cone. The presence of Rashba SOC has turned the Dirac cone dispersion into parabolic bands, separated by amount $\lambda$ in the middle region. The solid black circles with labels A-F mark the different states that an electron with spin-up (A) at chemical potential $\mu$, hitting the interface at $x=0$, can experience during the scattering process. The horizontal blue arrows and vertical black arrows show the particle's propagation and spin directions, respectively.  }
\end{figure*}
\begin{subequations}
\begin{align}
r_{\downarrow\downarrow}=& +\frac{2i}{D} e^{i \theta_\downarrow}(\sin \theta_\downarrow-\sin\theta_\uparrow )\sin^2\frac{\alpha}{2},\\
r_{\downarrow\uparrow}=& +\frac{e^{-i\gamma}}{D}   (e^{i\theta_\downarrow} - e^{i\theta_\uparrow}) \sqrt{\cos\theta_\downarrow \cos\theta_\uparrow}\sin\alpha,\\
t_{\downarrow\downarrow}=& +\frac{4}{D}\cos\theta_\downarrow \cos\theta_\uparrow\cos\frac{\alpha}{2},\\
t_{\downarrow\uparrow}=& -\frac{4e^{-i\gamma}}{D}e^{i\frac{\theta_\downarrow-\theta_\uparrow}{2}}\sqrt{\cos\theta_\downarrow\cos\theta_\uparrow}\cos\frac{\theta_\uparrow+\theta_\downarrow}{2}\sin\frac{\alpha}{2}.
\end{align}
\end{subequations}
Replacing these coefficients into Eqs.~(\ref{jx})-(\ref{jz}) and assuming $\theta_\uparrow\neq \theta_\downarrow\ll 1$, we find the following expressions for the components of STT
\begin{subequations}
\begin{align}\label{t_x}
&\boldsymbol{T}_x = \frac{\theta_\uparrow\theta_\downarrow\sin\alpha}{16}\Big(\theta_\uparrow\theta_\downarrow(\rho_\downarrow-\rho_\uparrow)(5\cos\alpha -1)\cos\gamma  \nonumber \\
&-2(\rho_\downarrow+\rho_\uparrow)(\theta_\uparrow+\theta_\downarrow)\sin^2\frac{\alpha}{2}\sin\gamma   \Big), 
\end{align}
\begin{align}\label{t_y}
&\boldsymbol{T}_y = \frac{\theta_\uparrow\theta_\downarrow\sin\alpha}{16}\Big(\theta_\uparrow\theta_\downarrow(\rho_\downarrow-\rho_\uparrow)(5\cos\alpha -1)\sin\gamma  \nonumber \\
&+2(\rho_\downarrow+\rho_\uparrow)(\theta_\uparrow+\theta_\downarrow)\sin^2\frac{\alpha}{2}\cos\gamma   \Big), 
\end{align}
\begin{align}\label{t_z}
&\boldsymbol{T}_z =\frac{\rho_\downarrow-\rho_\uparrow}{4} \Big( \theta_\downarrow^2 + (1+\theta_\downarrow^2)\theta_\uparrow^2 + (4\theta_\uparrow\theta_\downarrow - \nonumber \\ &\theta_\uparrow^2 +\theta_\downarrow^2(\theta_\uparrow^2-1) )\cos\alpha+ \theta_\uparrow^2\theta_\downarrow^2\cos 2\alpha \Big). 
\end{align}
\label{txyz}
\end{subequations}
The results illustrate that $\boldsymbol{T}_x$ and $\boldsymbol{T}_y$ are similar except a $\pi/2$ shift in the azimuthal angle $\gamma$. Namely, to obtain $\boldsymbol{T}_y$, Eq.~(\ref{t_y}), one needs to replace $\gamma$ by $\gamma+\pi/2$ in the expression for $\boldsymbol{T}_x$, Eq.~(\ref{t_x}). As is seen in Eq.~(\ref{t_z}), the z component of STT depends solely on the polar angle $\alpha$ and is independent of the azimuthal angle $\gamma$.   
Interesting phenomena appear when the incoming fermions propagate perpendicularly to the interfaces. To gain insight into this regime, one simply sets $\theta_\uparrow=\theta_\downarrow=0$. In this limit, using Eqs.~(\ref{Eq.FF.RandT}), we obtain $r_{\uparrow \uparrow}=r_{\uparrow \downarrow}=r_{\downarrow\uparrow}=r_{\downarrow\downarrow}=0$ and
\begin{subequations}
\begin{align}
t_{\uparrow \uparrow}=&+\cos\frac{\alpha}{2},\\
t_{\uparrow \downarrow}=&+\sin\frac{\alpha}{2}e^{+i\gamma},\\
t_{\downarrow \downarrow}=&+\cos\frac{\alpha}{2},\\
t_{\downarrow \uparrow}=&-\sin\frac{\alpha}{2}e^{-i\gamma}.
\end{align}
\label{Eq.Klein1}
\end{subequations}
Therefore, regardless of the spin orientation, Klein tunneling \cite{Katsnelson2006NP,Solnyshkov2016PRB,Liu2012PRB} occurs for these incoming spin-polarized fermions so that these fermions pass through the junction without any back scattering. Also, as can be seen in Eqs.~(\ref{Eq.Klein1}), these fermions can change their spin-channel, during the process of transmitting from the left ferromagnetic region into the right ferromagnetic region, upon the rotation of magnetization in the right ferromagnetic region. By substituting Eqs.~(\ref{Eq.Klein1}) into Eqs.~(\ref{jx})-(\ref{jz}) or simply setting $\theta_\uparrow=\theta_\downarrow=0$ in Eqs.~(\ref{txyz}), we interestingly find that $\boldsymbol{T}_x=\boldsymbol{T}_y=\boldsymbol{T}_z=0$ despite the fact that the spin-polarized Dirac fermions can change their spin-channel, Eqs.~(\ref{Eq.Klein1}). Therefore, one can conclude that the spin-polarized relativistic fermions in this situation are unable to produce STT and magnetoresistance and travel throughout the sample independent of spin rotation. Note that this picture is valid for those Dirac fermions that hit the interfaces at perpendicular angle, even in the presence of a barrier at the interfaces, and specific to the relativistic Dirac fermions. Our further calculations (not shown) illustrate that the presence of a thick enough SOC region limits the Klein phenomenon and the spin-polarized Dirac fermions with perpendicular trajectory to the interfaces exert STT upon the misalignment of the magnetization orientations in the two ferromagnetic regions. The physical reason behind this striking difference resides in the modification of the band structure of graphene by SOC. In the presence of SOC, the linear dispersion of Dirac fermions turns into a parabolic and thereby slows down the relativistic Dirac fermions. Also, this effect is absent in a conventional F-F junction where quasiparticles are governed by a parabolic dispersion relation and quasiparticles that hit the interface perpendicular to the interface exert nonzero STT to the junction. In the following, we shall study the spin-valve configuration with SOC (Fig.~\ref{fig1}) where all possible trajectories and propagation angles are accounted. 

\begin{figure}[t!]
\includegraphics[width=0.45\textwidth]{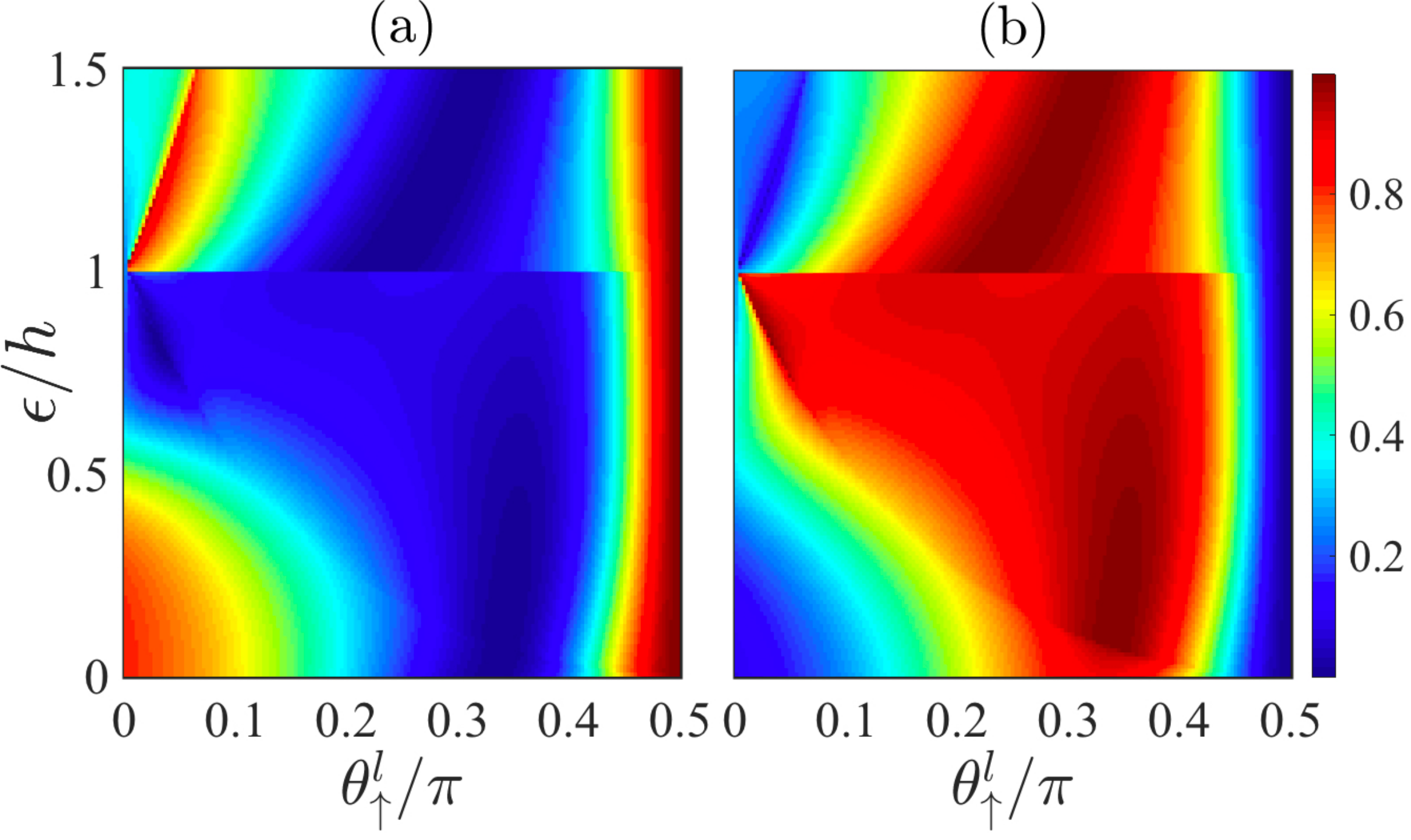}
\caption{\label{ruu_tuu} (Color online). (a) Reflection $|r_{\uparrow\uparrow}|$ and (b) transmission $|t_{\uparrow\uparrow}|$ vs the energy of spin up quasiparticle $\epsilon/h$ and the associated incident angle $\theta^l_\uparrow$ in the left F region. The remaining parameters are set identical to those used in the bottom row panels of Fig. \ref{stt_mu} with $\alpha=\gamma=0.75\pi$. }
\end{figure}
\begin{figure*}[th!]
\includegraphics[width=0.9\textwidth]{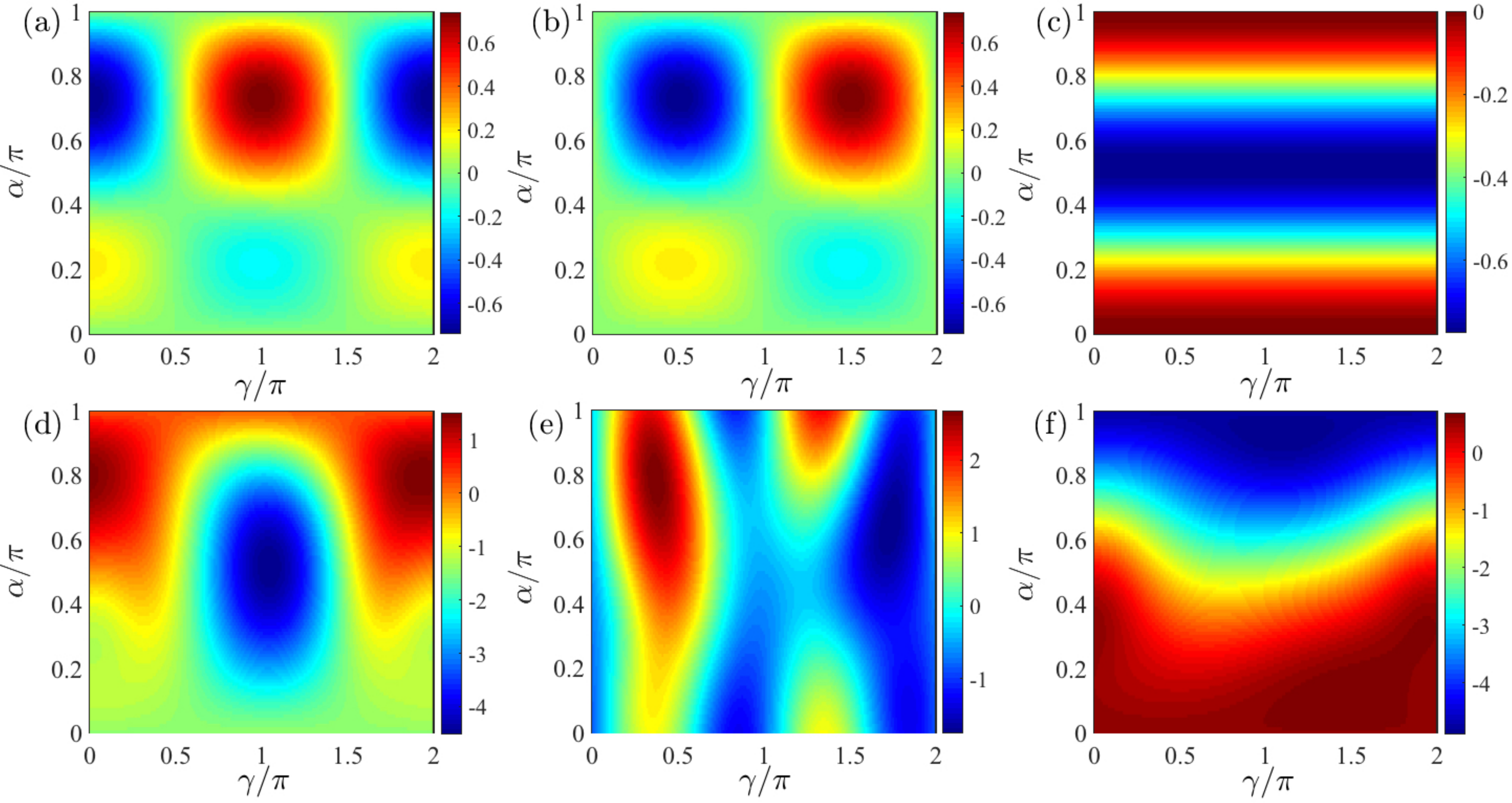}
\caption{\label{2d_stt} (Color online). The STT as a function of $\alpha$ and $\gamma$, the polar and azimuthal angles of magnetization in the right ferromagnetic region. In (a)-(c) no SOC is present whereas in (d)-(e) SOC has a finite representative value $\lambda=h$. The left, middle, and right columns show the $x$, $y$, and $z$ components of STT. The chemical potential and junction length are set to $\mu=1.5h$ and $d=\ell_F$, respectively. }
\end{figure*} 

In Fig.~\ref{stt_mu}, STT is plotted as a function of chemical potential, $\mu$. The magnetization in the left ferromagnetic region of Fig.~\ref{fig1} is fixed along the $z$ axis and the polar angle of the right ferromagnetic region in Fig.~\ref{fig1} is set to a representative value, i.e., $\alpha=0.75\pi$. To study how the magnetization misalignment alters STT, we vary the azimuthal angle, i.e., $\gamma=0.0, 0.25\pi, 0.5\pi, 0.75\pi, \pi$ in the right F region. The junction length is fixed at $d=2\ell_F$. In Figs.~\ref{stt_mu}(a)-\ref{stt_mu}(c), SOC is absent, i.e., $\lambda=0$ while in Figs.~\ref{stt_mu}(d)- \ref{stt_mu}(f), SOC has a finite and representative value, i.e., $\lambda=1.5h$. As is seen in Figs.~\ref{stt_mu}(a) and \ref{stt_mu}(b), $\boldsymbol{T}_x$ and $\boldsymbol{T}_y$ oscillate against $\mu$ whereas $\boldsymbol{T}_z$ responds monotonically with little oscillations with respect to the increase of $\mu$. The enhanced amplitudes can be understood by noting the fact that the introduced increase in $\mu$ results in increasing the population of particles, and therefore, enhancing the quasiparticle current. By increasing $\gamma$, the amplitude of the oscillations in $\boldsymbol{T}_{x,y}$ enhances and acquires a shift towards larger values of $\mu$. As is apparent in Fig.~\ref{stt_mu}(c), the $\boldsymbol{T}_{z}$ component of STT is insensitive to the azimuthal angle $\gamma$. The behavior of $\boldsymbol{T}_{x,y,z}$ shall be discussed further below. Note that the little abrupt change seen in Figs.~\ref{stt_mu} appears when the chemical potential crosses the Dirac point of the spin down subband and therefore the contribution of this subband is eliminated at $\mu=h$. To further elaborate on this process, we have shown the scattering process of a quasiparticle with spin up in Fig.~\ref{bs}. This particle in state A, belong to the spin up subband, within the left F region, propagates to the right along the $x$ axis and hits the interface at $x=0$. The presence of SOC in the middle region can flip the spin direction so that the reflected or transmitted particle can pass through F or C states, respectively. In this scenario, the increase of chemical potential $\mu$ shifts the energy of quasiparticle and at $\mu=h$, the spin down subband is unable to contribute to the scattering process. This switching from the conduction band to valance band appears as a small jump in the reflection and transmission coefficients at states E and D, respectively. Figure~\ref{ruu_tuu} displays the reflection and transmission probabilities when a spin up particles at state A (Fig.~\ref{bs}) is scattered through states E and D, upon hitting the junction interface at $x=0$, as a function of quasiparticle's energy $\epsilon/h$ and incident angle $\theta^l_\uparrow$. The junction length, strength of SOC, and the azimuthal and polar angles are set at $d=2\ell_F, \lambda=1.5h, \gamma=0.75\pi$, and $\alpha=0.75\pi$, respectively, corresponding to those of the bottom row of Fig.~\ref{stt_mu}. As is clearly seen, the probabilities show an abrupt change at $\epsilon=h$ where the contribution of spin down subband vanishes and the quasiparticle's energy crosses the Dirac point of the spin down subband in the F region. Note that this abrupt change is independent of $\lambda$ and depends on the magnetization misalignment. Therefore, it disappears at $\alpha=\gamma=0$ where the spin mixed backscattering and transmission probabilities are zero and practically the contribution of the spin down subband in the scattering process vanishes.         

\begin{figure*}[t!]
\includegraphics[width=\textwidth]{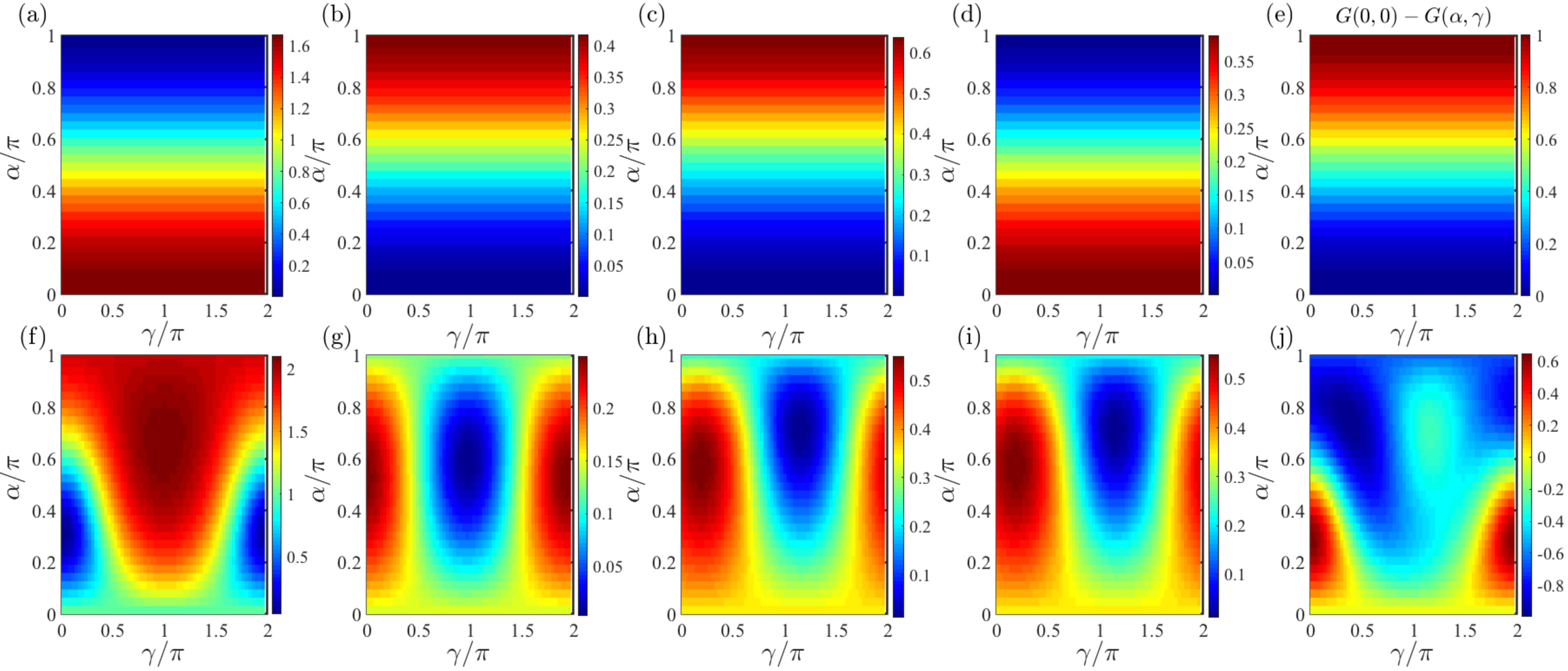}
\caption{\label{2d_GG} (Color online). Transmission probabilities $|t_{\uparrow\uparrow}|$, $|t_{\uparrow\downarrow}|$, $|t_{\downarrow\uparrow}|$, $|t_{\downarrow\downarrow}|$ and magnetoresistance $G(0,0)-G(\alpha,\gamma)$. In (a)-(e) we set $\lambda=0$ whereas in (f)-(j) the strength of SOC is set to $\lambda=h$. The chemical potential and junction length are $\mu=1.5h$ and $d=\ell_F$, respectively.}
\end{figure*}  
Figures \ref{stt_mu}(d)-\ref{stt_mu}(f) illustrate the behavior of the components $\boldsymbol{\cal T}_{x,y,z}$ of STT, Eq.~(\ref{Eq.STT1}), against chemical potential $\mu$. Due to the additional torque that SOC, Eq.~(\ref{Eq.STT1}), exerts, the total STT responds drastically differently to the increase of chemical potential compared to the previous case (Figs.\ref{stt_mu}(a)-\ref{stt_mu}(c)). By increasing $\gamma$ from $0$ to $\pi$, the component $\boldsymbol{\cal T}_{x}$ becomes smoother and is shifted downwards. Although $\boldsymbol{\cal T}_{y}$ shows smoother behavior when increasing $\gamma$ from $0$ to $\pi$, it first reaches a maximum at $\gamma=0.5\pi$ and then decreases. The $\boldsymbol{\cal T}_{z}$ component in the presence of SOC is modified anisotropically as a function $\gamma$ shown in Fig.~\ref{stt_mu}(f), which is in stark contrast to Fig.~\ref{stt_mu}(c) with no SOC.

To gain deeper insight, we plot the three components of STT as a function of $\alpha$ and $\gamma$ in Fig.~\ref{2d_stt}. The top and bottom row panels of Fig.~\ref{2d_stt} show $\boldsymbol{T}_{x,y,z}$ and $\boldsymbol{\cal T}_{x,y,z}$, respectively. The junction length, chemical potential, and strength of SOC are set to $d=\ell_F$, $\mu=1.5h$, and $\lambda=h$, respectively. Figures \ref{2d_stt}(a)-\ref{2d_stt}(c) illustrate the oscillatory behavior of STT as a function of $\alpha$. The component $\boldsymbol{T}_{z}$ is independent of $\gamma$ whereas $\boldsymbol{T}_{x,y}$ oscillate as a function of $\gamma$. These findings can be understood by considering Eq.~(\ref{Eq.STTSpinCurrent}). When magnetization in the two ferromagnetic regions of Fig.~\ref{fig1} are aligned, no spin torque appears since the cross product in Eqs.~(\ref{Eq.6b}) and (\ref{Eq.6c}) is zero. The maximum of STT along the $z$ axis, seen in Fig.~\ref{2d_stt}(c), appears when magnetization in the two ferromagnetic regions are perpendicular. The maximum of STT along the $x$ and $y$ spin axes possess a $\pi/2$ shift as a function of $\gamma$. The maximum values of $\boldsymbol{T}_{x,y}$ appear at $\alpha=\pi/4,3\pi/4$. Note that the numerical results are in agreement with the analytical expressions (Eqs.~(\ref{txyz})) obtained for a low-angle trajectory regime, i.e., $\theta_{\uparrow},\theta_{\downarrow}\ll 1$. In Figs.~\ref{2d_stt}(d)-\ref{2d_stt}(f), we set a nonzero value to the strength of SOC, i.e., $\lambda=h$. We clearly observe drastic changes to the components of STT. The maximum values of $\boldsymbol{T}_{x}$, illustrated in Fig.~\ref{2d_stt}(d), occur at $\alpha\approx \pi/2, 3\pi/4$ with nonzero values at $\gamma=0$ and $\gamma=\pi$ in contrast to the case when $\lambda=0$. The anisotropic modification introduced by SOC causes similar rearrangement of the maximum values of $\boldsymbol{T}_{y}$ in $\alpha$-$\gamma$ space as shown in the colormap plot Fig.~\ref{2d_stt}(e). As is seen in Fig.~\ref{2d_stt}(f), the maximum of $\boldsymbol{T}_{z}$ occurs with oscillations around $\alpha=\pi$ as a function of $\gamma$. Therefore, by subtracting the total STT in the presence and absence of SOC, one is able to extract and isolate the contribution of STT due to the induced SOC into the graphene system.

One of the most accessible quantities in experiment is charge conductance and associated magnetoresistance in magnetic junctions. \cite{C.Shen,Zhou,M.Diez,J.Moser,prox10,stronics4,A.M.A1,A.M.A2,MZare} This quantity can be obtained by following the same derivation steps described in deriving Eqs.~(\ref{jx})-(\ref{jz}) except now for charge current. After performing these calculations, we arrive at 
\begin{equation}
G(\alpha,\gamma)= \sum_{\delta,\delta'=\uparrow,\downarrow} G_{0,\delta} \int |t_{\delta,\delta'}|^2\cos\theta_{\delta'}d\theta_{\delta'}.
\end{equation}
The charge conductance is compromised of the transmission probabilities, i.e., $t_{\delta,\delta'}$. In Fig.~\ref{2d_GG}, we plot these probabilities as a function of $\alpha$ and $\gamma$. In Figs.~\ref{2d_GG}(a)-\ref{2d_GG}(d) and Figs.~\ref{2d_GG}(f)-\ref{2d_GG}(i), the strength of SOC is set to $\lambda=0$ and $\lambda=h$, respectively. This is evident that SOC yields an anisotropic charge conductance as a function of $\gamma$. In Figs.~\ref{2d_GG}(e) and \ref{2d_GG}(j), magnetoresistance, i.e., $G(0,0)-G(\alpha,\gamma)$ is plotted, which shows an anisotropically modified behavior in the presence of SOC.
 
We note that the maxima/minima shown in Figs.~\ref{stt_mu}-\ref{2d_GG} in STT, charge conductance, and magnetoresistance change locations by varying the junction thickness. This occurs because of the interference of the spin-polarized Dirac fermions bouncing back and forth between the two interfaces. Due to the different populations of spin-up and spin-down particles and band spin splittings, the interference of quasiparticles moving in the spin-up and spin-down channels is incommensurate, and therefore the final interference pattern is a superposition of interference in these two channels.  
Our findings clearly illustrate that one is able to detect and evaluate SOC in a proximity graphene device by STT, charge conductance, and magnetoresistance. \cite{D.Wang,A.Dyrdat,K.-H.Ding,B.Zhou,J.Chen,F.Herling,C.-C.Lin2,N.Tombros,Z.P.Niu,C.-C.Lin}

\section{conclusions}\label{conclusion}

In conclusion, using a generalized formulation for a graphene layer with spin orbit coupling (SOC) and magnetization, we have studied spin transfer torque (STT), charge conductance, and magnetoresistance in a spin orbit mediated spin-valve configuration. Our results demonstrate that the presence of SOC in a graphene layer results in anisotropically modifed STT, charge conductance, and magnetoresistance as a function of the relative magnetization misalignment. These quantities however show isotropic responses to the relative magnetization misalignment in the two ferromagnetic regions that allows for detecting the induced SOC in graphene. Interestingly, in the Klein regime where quasiparticles hit the interfaces at perpendicular direction, our results illustrate that the spin-polarized Dirac fermions move through a ferromagnet-ferromagnet graphene junction without experiencing any spin torque, regardless of the orientation of magnetization in the ferromagnetic regions.

\end{document}